\documentclass[draftcls,onecolumn]{IEEEtran}

\usepackage[cmex10]{amsmath}
\usepackage{amssymb}
\usepackage{graphicx}
\usepackage{epstopdf}
\usepackage{url}

\newcommand{\alphaAl}{\alpha_{{\rm A},l}}
\newcommand{\alphaAlstar}{\alpha_{{\rm A},l^*}}
\newcommand{\alphaBl}{\alpha_{{\rm B},l}}
\newcommand{\alphaBlstar}{\alpha_{{\rm B},l^*}}
\newcommand{\argmax}{\operatornamewithlimits{argmax}}
\newcommand{\EoneDDFNC}{E_1^{\rm DDF\&NC}}
\newcommand{\EoneEDDFNC}{E_1^{\rm EDDF\&NC}}
\newcommand{\EtwoDDFNC}{E_2^{\rm DDF\&NC}}
\newcommand{\EtwoEDDFNC}{E_2^{\rm EDDF\&NC}}

\newcommand{\ElstarADDFNC}{E_{l^*,{\rm A}}^{\rm DDF\&NC}}
\newcommand{\ElstarAEDDFNC}{E_{l^*,{\rm A}}^{\rm EDDF\&NC}}

\newcommand{\ElstarBDDFNC}{E_{l^*,{\rm B}}^{\rm DDF\&NC}}
\newcommand{\ElstarBEDDFNC}{E_{l^*,{\rm B}}^{\rm EDDF\&NC}}
\newcommand{\hAB}{h_{{\rm A},{\rm B}}}
\newcommand{\hABs}{|h_{{\rm A},{\rm B}}|^2}
\newcommand{\hBA}{h_{{\rm B},{\rm A}}}
\newcommand{\hAls}{|h_{{\rm A},l}|^2}
\newcommand{\hAlstar}{h_{{\rm A},l^*}}
\newcommand{\hAlstars}{|\hAlstar|^2}
\newcommand{\hBls}{|h_{{\rm B},l}|^2}
\newcommand{\hBlstar}{h_{{\rm B},l^*}}
\newcommand{\hBlstars}{|\hBlstar|^2}
\newcommand{\hlstarB}{h_{l^*,{\rm B}}}
\newcommand{\hlstarBs}{|h_{l^*,{\rm B}}|^2}
\newcommand{\hlstarA}{h_{l^*,{\rm A}}}
\newcommand{\hlstarAs}{|h_{l^*,{\rm A}}|^2}
\newcommand{\hXY}{h_{X,Y}}
\newcommand{\JprimeAlstar}{J'_{{\rm A},l^*}}
\newcommand{\JprimeBlstar}{J'_{{\rm B},l^*}}

\newcommand{\OABDDFNC}{O_{\rm A,B}^{\rm DDF\&NC}}
\newcommand{\OABEDDFNC}{O_{\rm A,B}^{\rm EDDF\&NC}}
\newcommand{\OADDFNC}{O_{\rm A}^{\rm DDF\&NC}}
\newcommand{\OAEDDFNC}{O_{\rm A}^{\rm EDDF\&NC}}

\newcommand{\OBADDFNC}{O_{\rm B,A}^{\rm DDF\&NC}}
\newcommand{\OBAEDDFNC}{O_{\rm B,A}^{\rm EDDF\&NC}}
\newcommand{\OBDDFNC}{O_{\rm B}^{\rm DDF\&NC}}
\newcommand{\OBEDDFNC}{O_{\rm B}^{\rm EDDF\&NC}}

\newcommand{\OlstarADDFNC}{O_{l^*,{\rm A}}^{\rm DDF\&NC}}
\newcommand{\OlstarAEDDFNC}{O_{l^*,{\rm A}}^{\rm EDDF\&NC}}

\newcommand{\OlstarBDDFNC}{O_{l^*,{\rm B}}^{\rm DDF\&NC}}
\newcommand{\OlstarBEDDFNC}{O_{l^*,{\rm B}}^{\rm EDDF\&NC}}
\newcommand{\ORSDDFNC}{O^{\rm RS-DDF\&NC}}
\newcommand{\ORSEDDFNC}{O^{\rm RS-EDDF\&NC}}
\renewcommand{\P}{{\rm P}}
\newcommand{\POABDDFNC}{\P[\OABDDFNC]}
\newcommand{\Pout}{P_{\rm out}}
\newcommand{\PoutRSDDFNC}{\Pout^{\rm RS-DDF\&NC}}
\newcommand{\PoutRSEDDFNC}{\Pout^{\rm RS-EDDF\&NC}}
\newcommand{\rhoAl}{\rho_{{\rm A},l}}
\newcommand{\rhoAlstar}{\rho_{{\rm A},l^*}}
\newcommand{\rhoBl}{\rho_{{\rm B},l}}
\newcommand{\rhoBlstar}{\rho_{{\rm B},l^*}}
\newcommand{\SNR}{{\rm SNR}}
\newcommand{\SNRnorm}{{\rm SNR}_{\rm norm}}
\newcommand{\Srelay}{S_{\rm relay}}
\newcommand{\wAn}{w_{\rm A}[n]}
\newcommand{\wBn}{w_{\rm B}[n]}
\newcommand{\wlstarn}{w_{l^*}[n]}

\newcommand{\yAn}{y_{\rm A}[n]}
\newcommand{\yBn}{y_{\rm B}[n]}
\newcommand{\ylstarn}{y_{l^*}[n]}

\newtheorem{lemma}{Lemma}
\newtheorem{theorem}{Theorem}

\begin{document}

\title{Performance Analysis of Relay Selection With Enhanced Dynamic Decode-and-Forward and Network Coding in Two-Way Relay Channels}

\author{Wei-Cheng~Liu~\IEEEmembership{Member,~IEEE} and Yu-Chen~Liu
\thanks{Manuscript received January 31, 2016. This work is supported by the National Science Council, Taiwan, under the contract NSC 101-2221-E-194-037.}
\thanks{W.-C.~Liu is with the Department of Communications Engineering, National Chung Cheng University, Chia-Yi, Taiwan (e-mail: comwcliu@ccu.edu.tw).}
\thanks{Y.-C.~Liu is with the Graduate Institute of Communication Engineering, National Taiwan University, Taipei, Taiwan (e-mail: ternence.cm96@g2.nctu.edu.tw).}}

\maketitle

\begin{abstract}
In this paper, we adopt the relay selection (RS) protocol proposed by Bletsas, Khisti, Reed and Lippman (2006) with Enhanced Dynamic Decode-and-Forward (EDDF) and network coding (NC) system in a two-hop two-way multi-relay network. All nodes are single-input single-output (SISO) and half-duplex, i.e., they cannot transmit and receive data simultaneously. The outage probability is analyzed and we show comparisons of outage probability with various scenarios under Rayleigh fading channel. Our results show that the relay selection with EDDF and network coding (RS-EDDF\&NC) scheme has the best performance in the sense of outage probability upon the considered decode-and-forward (DF) relaying if there exist sufficiently relays. In addition, the performance loss is large if we select a relay at random. This shows the importance of relay selection strategies.
\end{abstract}


\section{Introduction}
\label{sec:Introduction}

In recent years, cooperative communications have attracted considerable interests and studies in communications. In cooperative communication networks, relay
terminals share their antennas and other resources to create a “virtual array” through distributed transmission and signal processing to against the fading arising from multipath propagation in wireless communications. Thus, cooperative communication systems can be viewed as virtual multi-input multi-output (MIMO) systems and provide a space diversity. This form of space diversity is called cooperative diversity (cf., user cooperation diversity of \cite{Sendonaris1998}). Hence, the diversity-multiplexing tradeoff (DMT) \cite{Zheng2003} framework can be adopted to the cooperative communication systems for further perspective. Many well-known half-duplex cooperative protocols with single relay have been identified in the literature such as amplify-and-forward (AF), decode-and-forward (DF), selection DF (SDF), incremental AF (IAF) \cite{Laneman2004}, distributed space-time coding (DSTC) \cite{Laneman2003}, nonorthogonal AF (NAF), dynamic DF (DDF) \cite{Azarian2005}, enhanced static DF (ESDF), enhanced DDF (EDDF) \cite{Prasad2010} and cooperative network coding (CNC) \cite{Wang2011}. Moreover, the cooperative communications can extend many advanced topics such as beamforming \cite{Fertl2008}, opportunistic relaying \cite{Bletsas2006JSAC}, cooperative relaying in orthogonal frequency-division multiplexing (OFDM) \cite{Han2009} and MIMO \cite{Munoz2005} systems. There are still many areas to be explored in cooperative communications.

In multi-relay cooperative communication systems, relays are usually assumed to transmit over orthogonal channels in order to avoid the interference. These orthogonal cooperation schemes result in low bandwidth efficiency as the number of relays increase. To overcome this problem, there are various cooperative communications schemes which have been developed such as beamforming, opportunistic relaying and DSTC.

\subsection{Related Work}
\label{sec:Related Work}

In \cite{Bletsas2006JSAC}, the authors proposed an opportunistic relaying protocol in which relays overhear the request-to-send (RTS) and clear-to-send (CTS) packets exchanged between the source and the destination before the data transmission. Then the optimal relay is selected by the selection criterion according to the quality of source-relay and relay-destination links. This protocol can be easily implemented in a decentralized way. It was proved that with the same intermediate relays, the DF based on this protocol achieves the same DMT of the DSTC scheme in \cite{Laneman2003}. In \cite{Bletsas2006WCNC}, the authors further considered reactive and proactive relay selection schemes. The authors proved that both reactive and proactive opportunistic DF relaying are outage-optimal. Moreover, the proactive strategy is much efficient since it can be viewed as energy-efficient routing in the network and reduces the overhead of the system. In \cite{Li2010}, the authors proposed two schemes: single relay selection with network coding (S-RS-NC) and dual relay selection with network coding (D-RS-NC), and show that the D-RS-NC scheme outperforms other considered RS schemes in two-way relay channels. More related studies about relay selection are listed in \cite{Bletsas2008}--\cite{Jing2009}.

\subsection{Goal}
\label{sec:Goal}

In our research, we would like to construct the system model of EDDF scheme and network coding based on the relay selection criterion of \cite{Bletsas2006JSAC} in a two-hop two-way multi-relay network, which is named RS-EDDF\&NC. We then analyze the outage probability of the system model and give numerical results of outage probability with various scenarios under Rayleigh fading channel.

\subsection{Paper Outline}
\label{sec:Paper Outline}

The rest of this paper is organized as follows. Section \ref{sec:RS-DDF&NC} gives the system model of relay selection with DDF and network coding, which is named RS-DDF\&NC. We also derive the outage probability of the RS-DDF\&NC system. In section \ref{sec:RS-EDDF&NC}, we show the system model and analyze the outage probability of the RS-EDDF\&NC scheme. In section \ref{sec:Numerical Results}, we give numerical results of outage probability with various scenarios under Rayleigh fading channel. Conclusions are given in section \ref{sec:Conclusions}.

\section{Relay Selection with Dynamic Decode-and-Forward and Network Coding System}
\label{sec:RS-DDF&NC}

\subsection{System Model}
\label{sec:System Model of RS-DDF&NC}

To construct our system model, we utilize a baseband-equivalent, discrete-time channel model. For any two nodes $X$ and $Y$, we denote the channel coefficient from node $X$ to node $Y$ as $h_{X,Y}$, where $h_{X,Y}$'s are independent and identically distributed (i.i.d.) random variables.

Given the transmitted signal $x_k[n]$, assuming that ${\rm E}\{|x_k[n]|^2\}=P$ for all $n$, where $k \in \{a_1,b_1,c_1\}, c_1=a_1\oplus b_1$. We denote $a_1$ and $b_1$ as information bit streams of user A and user B, respectively. The subscript `1' denotes that the information bit stream was encoded by using codebook 1.

Noise at each receiver is assumed to be i.i.d. circularly symmetric complex Gaussian random variables $CN(0,N_0)$. We denote the noise at the node $X$ during the $n$th symbol interval as $w_X[n]$. We will first list some main assumptions then describe each phase of the system model for detail in the following subsections.

\subsection{Assumptions}
\label{sec:Assumptions}

The main assumptions are as follows:
\begin{enumerate}
\item We consider a two-hop two-way relay network that consists of $L$ relays.
\item All nodes are single-input single-output (SISO) and half-duplex.
\item Codewords transmitted from user A and user B both have the same frame duration of $J$ symbol intervals.
\item The codeword transmitted by the source is of the \emph{incremental redundancy} type in which it can be decoded by observing just one or a few of its segments.
\item We assume all the channel coefficients remain constant during the whole cooperative transmission intervals. The forward and backward channels between two nodes $X$ and $Y$ are the same during the whole cooperative transmission due to the \emph{reciprocity theorem} \cite{Rappaport1996}, i.e., $h_{X,Y}=h_{Y,X}$.
\item Channel state information (CSI) is not available at the source (user A and user B).
\item CSI between the source to relay $l$ and relay $l$ to the destination is available at each relay $l$, $l=1, 2, ..., L$.
\item We assume the use of random Gaussian codebooks, and the length of the codeword is sufficiently long.
\item For relay selection, we assume all relays are hidden from each other and we ignore losses, delays and collisions of relays' \emph{flag} packets \cite{Bletsas2006JSAC} and destination's notification packet.
\end{enumerate}

\subsection{Phase 1 and Phase 2 of RS-DDF\&NC}
\label{sec:Phase 1 and Phase 2 of RS-DDF&NC}

\begin{figure}
\centering
\includegraphics{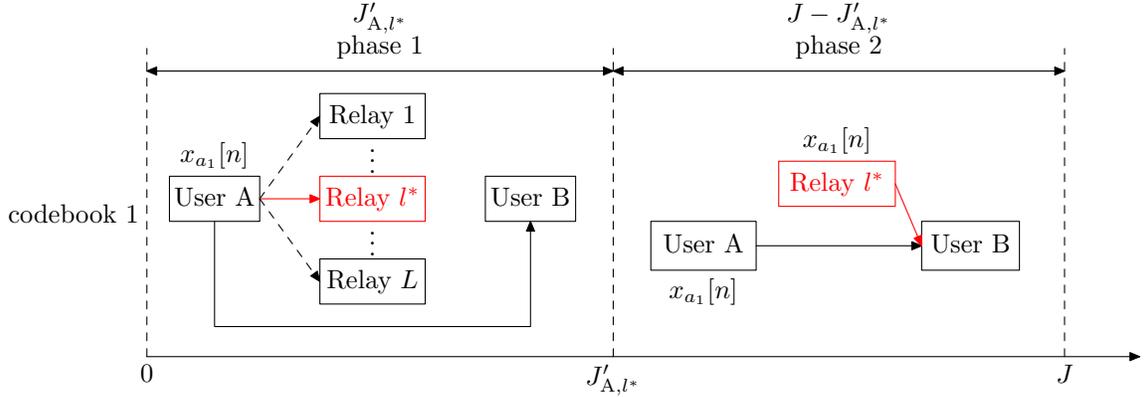}
\caption{Illustration of phase 1 and phase 2 for the RS-DDF\&NC scheme.}
\label{fig:RS_DDF_NC_phases_1_and_2}
\end{figure}

The phase 1 and phase 2 of the RS-DDF\&NC are shown in Fig. \ref{fig:RS_DDF_NC_phases_1_and_2}. Before the data transmission, the optimal relay $l^*$ was selected according to the relay selection criterion given by \cite{Bletsas2006JSAC}
\begin{equation}\label{eq:l^*}
l^*=\argmax_{l\in\Srelay} q_l,
\end{equation}
where
\begin{equation}\label{eq:q_l}
q_l=\min\{|h_{{\rm A},l}|^2,|h_{l,{\rm B}}|^2\}\quad \forall\ l\in\Srelay,
\end{equation}
and $\Srelay=\{1,2,...,L\}$. Other relays then enter an idle mode so that they do not process the signal from the source. In phase 1, the source (user A) transmits $x_{a_1}[n]$ to the destination (user B), and only the best relay $l^*$ receives this signal and try to decode it. The received signals at the destination (user B) and the relay $l^*$ during the $n$th symbol interval are expressed as
\begin{equation}\label{eq:y_B[n]1}
\yBn=\hAB x_{a_1}[n]+\wBn
\end{equation}
and
\begin{equation}\label{eq:y_l^*[n]1}
y_{l^*}[n]=\hAlstar x_{a_1}[n]+w_{l^*}[n],
\end{equation}
respectively, for $0<n\leq\JprimeAlstar$, where $\JprimeAlstar$ is defined as
\begin{equation}\label{eq:JprimeAlstar}
\JprimeAlstar \triangleq \min \left\{J, \left\lceil \frac{J R}{\log_2(1+\rhoAlstar |\hAlstar|^2)} \right\rceil\right\},
\end{equation}
where $R$ (bps/Hz) is the spectrum efficiency and $\rhoAlstar$ is the SNR of the link between user A and relay $l^*$.

In phase 2, relay $l^*$ re-encodes the message by using the same codebook if the outage did not occur. Relay $l^*$ then cooperatively transmits the corresponding part of the codeword to the destination (user B) in the remaining intervals. The received signals at the destination (user B) in this phase can be expressed as
\begin{equation}\label{eq:y_B[n]2}
\yBn=\hAB x_{a_1}[n]+\hlstarB x_{a_1}[n]+\wBn,\quad \text{for } \JprimeAlstar<n\leq J.
\end{equation}

\subsection{Phase 3 and Phase 4 of RS-DDF\&NC}
\label{sec:Phase 3 and Phase 4 of RS-DDF&NC}

\begin{figure}
\centering
\includegraphics{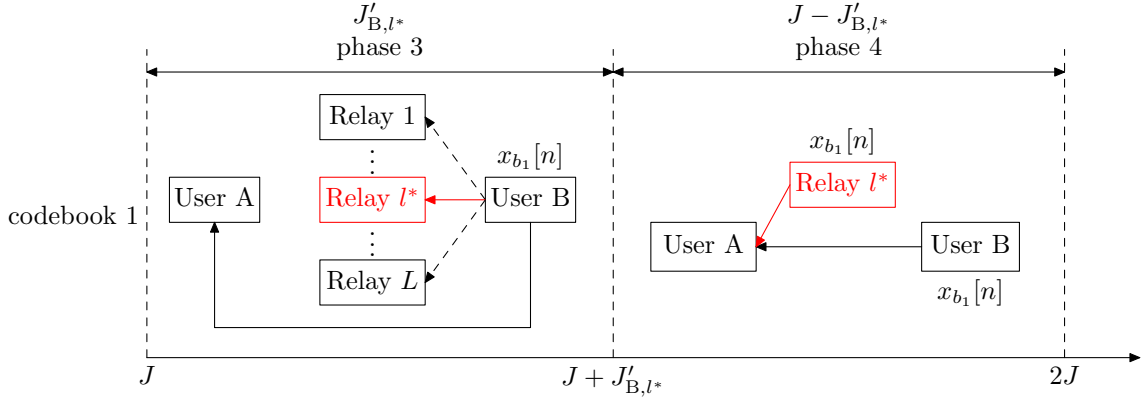}
\caption{Illustration of phase 3 and phase 4 for the RS-DDF\&NC scheme.}
\label{fig:RS_DDF_NC_phases_3_and_4}
\end{figure}

The phase 3 and phase 4 of the RS-DDF\&NC are shown in Fig. \ref{fig:RS_DDF_NC_phases_3_and_4}. Similar to phase 1 and phase 2, the source (user B) transmits $x_{b_1}[n]$ to the destination (user A), and only the best relay $l^*$ receives this signal. The received
signals at the destination (user A) and the relay $l^*$ during the $n$th symbol interval are denoted by $y_{\rm A}[n]$ and $y_{l^*}[n]$, respectively, expressed as
\begin{equation}\label{eq:y_A[n]3}
\yAn=\hBA x_{b_1}[n]+\wAn
\end{equation}
and
\begin{equation}\label{eq:y_l^*[n]3}
y_{l^*}[n]=\hBlstar x_{b_1}[n]+w_{l^*}[n],
\end{equation}
for $J<n\leq J+\JprimeBlstar$, where $\JprimeBlstar$ is defined as
\begin{equation}\label{eq:JprimeBlstar}
\JprimeBlstar \triangleq \min \left\{J, \left\lceil \frac{J R}{\log_2(1+\rhoBlstar |\hBlstar|^2)} \right\rceil\right\},
\end{equation}
where $\rhoBlstar$ is the SNR of the link between user B and relay $l^*$.

In phase 4, relay $l^*$ re-encodes the message by using the same codebook if the outage did not occur. Relay $l^*$ then cooperatively transmits the corresponding part of the codeword to the destination (user A) in the remaining intervals. The received signals at the destination (user A) in this phase can be expressed as
\begin{equation}\label{eq:y_A[n]4}
\yAn = \hBA x_{b_1}[n] + \hlstarA x_{b_1}[n] + \wAn,\quad \text{for } J + \JprimeBlstar < n \leq 2J.
\end{equation}

\subsection{Phase 5 of RS-DDF\&NC}
\label{sec:Phase 5 of RS-DDF&NC}

\begin{figure}
\centering
\includegraphics{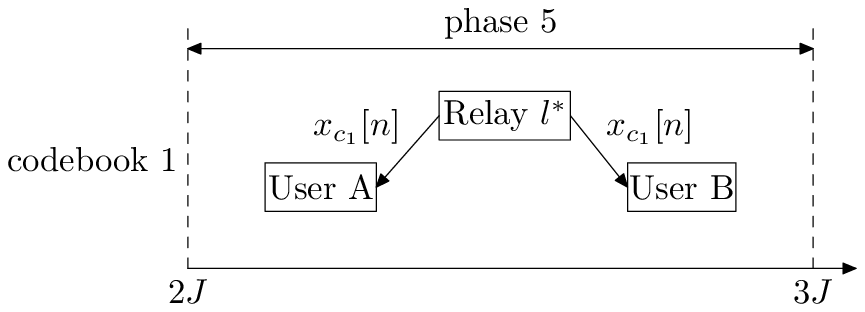}
\caption{Illustration of phase 5 for the RS-DDF\&NC scheme.}
\label{fig:RS_DDF_NC_phase_5}
\end{figure}

In phase 5, relay $l^*$ performs exclusive-or (XOR) on the two decoded messages from user A and user B and re-encodes it by using codebook 1. Relay $l^*$ then transmits this signal, denoted as $x_{c_1}[n]$, to both user A and user B. The received signals at user A and user B can be expressed as
\begin{equation}\label{eq:y_A[n]5}
\yAn=\hlstarA x_{c_1}[n]+\wAn
\end{equation}
and
\begin{equation}\label{eq:y_B[n]5}
\yBn=\hlstarB x_{c_1}[n]+\wBn,
\end{equation}
respectively, for $2J<n\leq3J$. The system model of phase 5 is shown in Fig. \ref{fig:RS_DDF_NC_phase_5}.

\subsection{Outage Probability of RS-DDF\&NC}
\label{sec:Outage Probability of RS-DDF&NC}

In formulating the outage events, we define the effective SNR as
\begin{equation}\label{eq:rho}
\rho \triangleq \frac{P}{N_0}.
\end{equation}
First, let us consider the outage event for user A. If outage events both occur on the link between user B and user A and the link between relay $l^*$ and user A, then user A will fail to decode the information of user B. Thus, we can express the outage event for user A as
\begin{equation}
\label{eq:OADDFNC}
\OADDFNC = \OBADDFNC \cap \OlstarADDFNC.
\end{equation}
Similarly, in the case for user B, user B fails to decode information of user A without error when outage events both occur on the link between user A and user B and the link between relay $l^*$ and user B. We express the outage event for user B as
\begin{equation}
\label{eq:OBDDFNC}
\OBDDFNC = \OABDDFNC \cap \OlstarBDDFNC.
\end{equation}

Next, we analyze outage events for links between user A, user B and relay $l^*$. The outage event between user A and user B is
\begin{align}
\label{eq:OABDDFNC}
\OABDDFNC
&=\left\{2J\log_2(1+\rho|\hAB|^2)<3JR\right\}\notag\\
&=\left\{\frac{2}{3}\log_2(1+\rho|\hAB|^2)<R\right\}.
\end{align}
Clearly, $\OBADDFNC = \OABDDFNC$ because $\hBA=\hAB$.

Moreover, for the outage event between relay $l^*$ and user A, it should be noted that relay $l^*$ has to decode the message from user B with arbitrarily small error in phase 3, or it will remain silent in phase 4 since it failed to decode the message. The outage event between relay $l^*$ and user A is
\begin{equation}
\label{eq:OlstarADDFNC}
\OlstarADDFNC = \EoneDDFNC \cup \ElstarADDFNC,
\end{equation}
where
\begin{align}
\label{eq:EoneDDFNC}
\EoneDDFNC
&=\left\{\JprimeBlstar=J\right\}\notag\\
&=\left\{\log_2(1+\rho|\hBlstar|^2)<\frac{JR}{J-1}\right\}
\end{align}
and
\begin{align}
\label{eq:ElstarADDFNC}
\ElstarADDFNC
&=\left\{(2J+\JprimeAlstar-\JprimeBlstar)\log_2(1+\rho|\hlstarA|^2)<3JR\right\}\notag\\
&=\left\{\frac{2J+\JprimeAlstar-\JprimeBlstar}{3J}\log_2(1+\rho|\hlstarA|^2)<R\right\}.
\end{align}

Similarly, the outage event between relay $l^*$ and user B is
\begin{equation}
\label{eq:OlstarBDDFNC}
\OlstarBDDFNC = \EtwoDDFNC \cup \ElstarBDDFNC,
\end{equation}
where
\begin{equation}
\label{eq:EtwoDDFNC}
\EtwoDDFNC=\left\{\log_2(1+\rho|\hAlstar|^2)<\frac{JR}{J-1}\right\}
\end{equation}
and
\begin{equation}
\label{eq:ElstarBDDFNC}
\ElstarBDDFNC=\left\{\frac{2J+\JprimeBlstar-\JprimeAlstar}{3J}\log_2(1+\rho|\hlstarB|^2)<R\right\}.
\end{equation}

The outage event of the overall system for RS-DDF\&NC can be written as
\begin{align}\label{eq:ORSDDFNC}
\ORSDDFNC
& = \OADDFNC \cup \OBDDFNC \notag \\
& = (\OBADDFNC \cap \OlstarADDFNC) \cup (\OABDDFNC \cap \OlstarBDDFNC) \notag \\
& = \OABDDFNC \cap (\OlstarADDFNC \cup \OlstarBDDFNC) \notag \\
& = \OABDDFNC \cap [(\EoneDDFNC \cup \ElstarADDFNC) \cup (\EtwoDDFNC \cup \ElstarBDDFNC)].
\end{align}
Let $Z_1 = \hAlstars$ and $Z_2 = \hBlstars$, the outage probability of RS-DDF\&NC can be expressed as
\begin{equation}
\label{eq:PoutRSDDFNC}
\PoutRSDDFNC = \int_0^\infty \int_0^\infty \P[\ORSDDFNC] f_{Z_1, Z_2}(z_1, z_2) dz_1 dz_2.
\end{equation}
In the following, we derive the outage probability of RS-DDF\&NC under Rayleigh fading scenario.
\begin{lemma}\label{lem:fZ1Z2}
Assume that channel coefficients $\hXY$ are i.i.d.~circularly symmetric complex Gaussian random variables with zero mean and unit variance. The joint probability density function (PDF) of $Z_1$ and $Z_2$ is
\begin{equation}\label{eq:fZ1Z2}
f_{Z_1, Z_2}(z_1, z_2)
\begin{cases}
L e^{-(z_1+z_2)} (1-e^{-2z_1})^{L-1}, & 0\leq z_1\leq z_2\\
L e^{-(z_1+z_2)} (1-e^{-2z_2})^{L-1}, & 0\leq z_2\leq z_1\\
0, & \textup{otherwise}. 
\end{cases}
\end{equation}
\end{lemma}

\begin{IEEEproof}
The proof is given in Appendix \ref{app:fZ1Z2}.
\end{IEEEproof}

The outage probability of RS-DDF\&NC scheme under Rayleigh fading scenario is analyzed in the following theorem:
\begin{theorem}\label{the:PoutRSDDFNC}
The outage probability of the RS-DDF\&NC scheme is
\begin{equation}\label{eq:PoutRSDDFNC2}
\PoutRSDDFNC=p_1 \int_0^\infty \int_0^\infty I(O) f_{Z_1,Z_2}(z_1,z_2) dz_1 dz_2,
\end{equation}
where
\begin{equation}\label{eq:p_1}
p_1=1-\exp\left(-\frac{2^{\frac{3R}{2}}-1}{\rho}\right),
\end{equation}
the event $O$ is defined in (\ref{eq:PEone...}), and
\begin{equation}
\label{eq:I(O)}
I(O)=
\begin{cases}
1, & \text{if } O \text{ occurs} \\
0, & \text{otherwise}.
\end{cases}
\end{equation}
The joint PDF $f_{Z_1,Z_2}(z_1,z_2)$ is given in Lemma \ref{lem:fZ1Z2}.
\end{theorem}

\begin{IEEEproof}
The proof is given in Appendix \ref{app:PoutRSDDFNC}.
\end{IEEEproof}

\section{Relay Selection with Enhanced Dynamic Decode-and-Forward and Network Coding System}
\label{sec:RS-EDDF&NC}

\subsection{System Model of RS-EDDF\&NC}
\label{sec:System Model of RS-EDDF&NC}

In the case of multiplexing gain $r>0.5$, the source (user A) transmits an additional independent codeword encoded by using another codebook (codebook 2) to the destination (user B) during intervals $(\beta J, J]$, and vice versa for intervals in $(J + \beta J, 2J]$. This can be achieved by employing superposition coding. In this case, codewords encoded by using codebook 1 are transmitted at a rate $R_1$ bps/Hz, and codewords
encoded by using codebook 2 are transmitted at a rate $R_2$ bps/Hz. These two transmission rates are determined by the EDDF protocol.

If multiplexing gain $r \leq 0.5$, the EDDF scheme is identical to the DDF scheme, with the optimal solution being $\hat{r}_1 = r$ and $\hat{\beta} = 1$, where $\hat{r}_1$ is the optimal multiplexing gain corresponding to $R_1$. The source only uses codebook 1 for transmission. In this case, the source transmits data at a rate $R$ bps/Hz during every symbol interval in the codeword.

The system model for the case $r \leq 0.5$ has been discussed in section \ref{sec:RS-DDF&NC}. Given the same system assumptions in the subsection \ref{sec:Assumptions}, we will describe the system model for each phase of RS-EDDF\&NC with multiplexing gain $r < 0.5$ in the following context.

\subsection{Phase 1 and Phase 2 of RS-EDDF\&NC}
\label{sec:Phase 1 and Phase 2 of RS-EDDF&NC}

\begin{figure}
\centering
\includegraphics{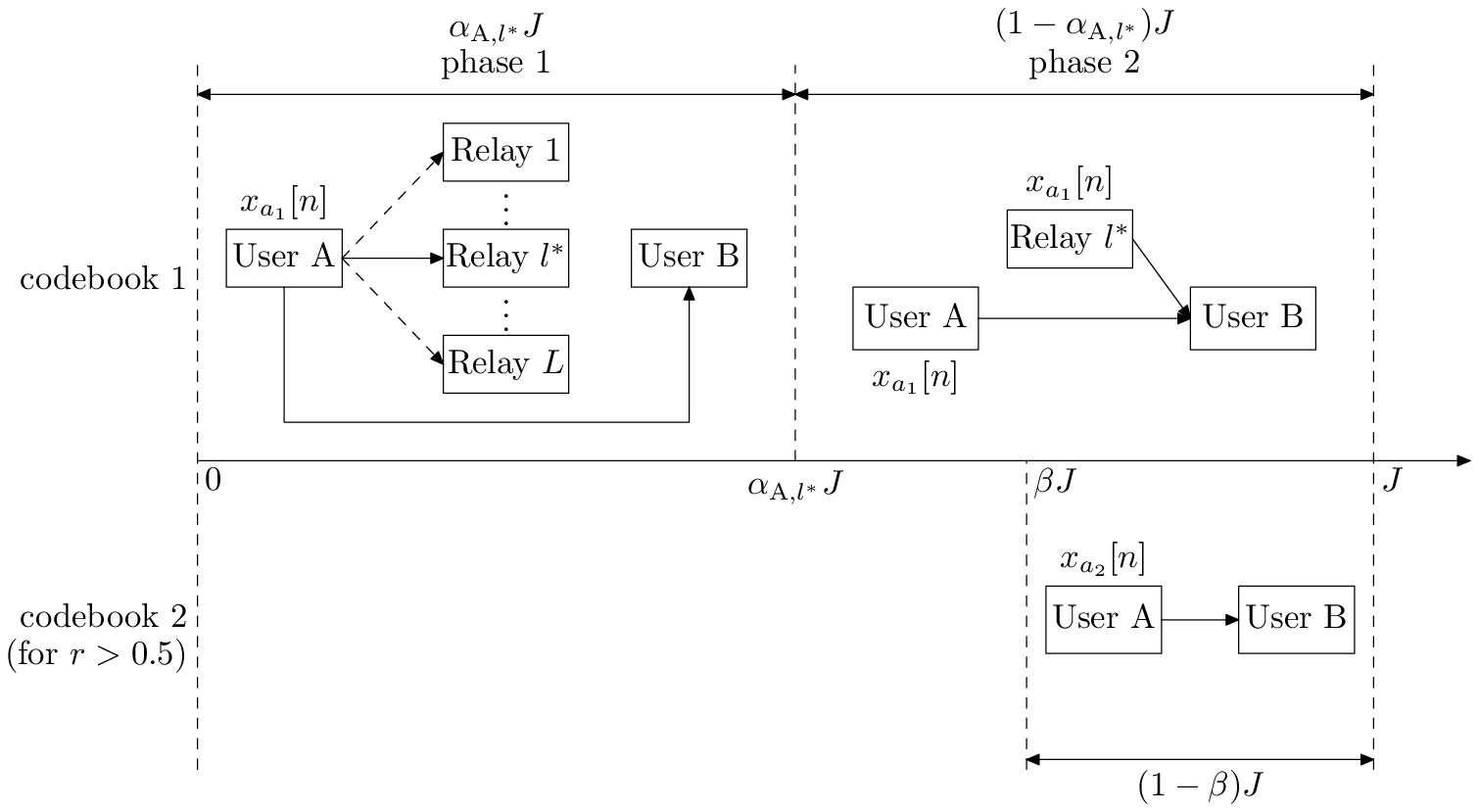}
\caption{Illustration of phase 1 and phase 2 for the RS-EDDF\&NC scheme.}
\label{fig:RS_EDDF_NC_phases_1_and_2}
\end{figure}

The first two phases (phase 1 and phase 2) of the RS-EDDF\&NC is shown in Fig.~\ref{fig:RS_EDDF_NC_phases_1_and_2}. The best relay $l^*$ was selected before the data transmission and other relays enter an idle mode. In phase 1, the source (user A) transmits $x_{a_1}[n]$ to the destination (user B) and the relay $l^*$ also receives this signal. The received signals at the destination (user B) and the relay $l^*$ during the $n$th symbol interval are expressed as
\begin{equation}
\yBn = \hAB x_{a_1}[n] + \wBn
\end{equation}
and
\begin{equation}
\ylstarn = \hAlstar x_{a_1}[n] + \wlstarn,
\end{equation}
respectively, for $0 < n \leq \alphaAlstar J$, where $\alphaAl$ is defined as
\begin{equation}
\alphaAl \triangleq \frac{r_1 \log_2(\rhoAl)}{\log_2(1 + \rhoAl \hAls)}, l = 1, ..., L,
\end{equation}
where $\rhoAl$ is the SNR from user A to relay $l$.

In phase 2, relay $l^*$ re-encodes the message by using the same codebook if the outage did not occur. Relay $l^*$ then cooperatively transmits the corresponding part of the codeword to the destination (user B) in the remaining intervals. In addition, the source (user A) transmits an additional independent codeword $x_{a_2}[n]$ using another codebook (codebook 2) to the destination (user B). The received signals at the
destination (user B) in this phase can be expressed as
\begin{equation}
\yBn =
\begin{cases}
\hAB x_{a_1}[n] + \hlstarB x_{a_1}[n] + \wBn, & \alphaAlstar J < n \leq \beta J \\
\hAB x_{a_1}[n] + \hlstarB x_{a_1}[n] + \hAB x_{a_2}[n] + \wBn, & \beta J < n \leq J,
\end{cases}
\end{equation}
where the optimal value for $\beta$ is given by Theorem 3 in \cite{Prasad2010}.

\subsection{Phase 3 and Phase 4 of RS-EDDF\&NC}
\label{sec:Phase 3 and Phase 4 of RS-EDDF&NC}

In phase 3, the source (user B) transmits $x_{b_1}[n]$ to the destination (user A), and the best relay $l^*$ receives this signal. The received signals at the destination (user A) and the relay $l^*$ during the $n$th symbol interval are denoted by $\yAn$ and $\ylstarn$, respectively,
expressed as
\begin{equation}
\yAn = \hBA x_{a_1}[n] + \wAn
\end{equation}
and
\begin{equation}
\ylstarn = \hBlstar x_{b_1}[n] + \wlstarn,
\end{equation}
for $J < n \leq J + \alphaBlstar J$, where $\alphaBl$ is defined as
\begin{equation}
\alphaBl \triangleq \frac{r_1 \log_2(\rhoBl)}{\log_2(1 + \rhoBl \hBls)}, l = 1, ..., L,
\end{equation}
where $\rhoBl$ is the SNR from user B to relay $l$.

\begin{figure}
\centering
\includegraphics{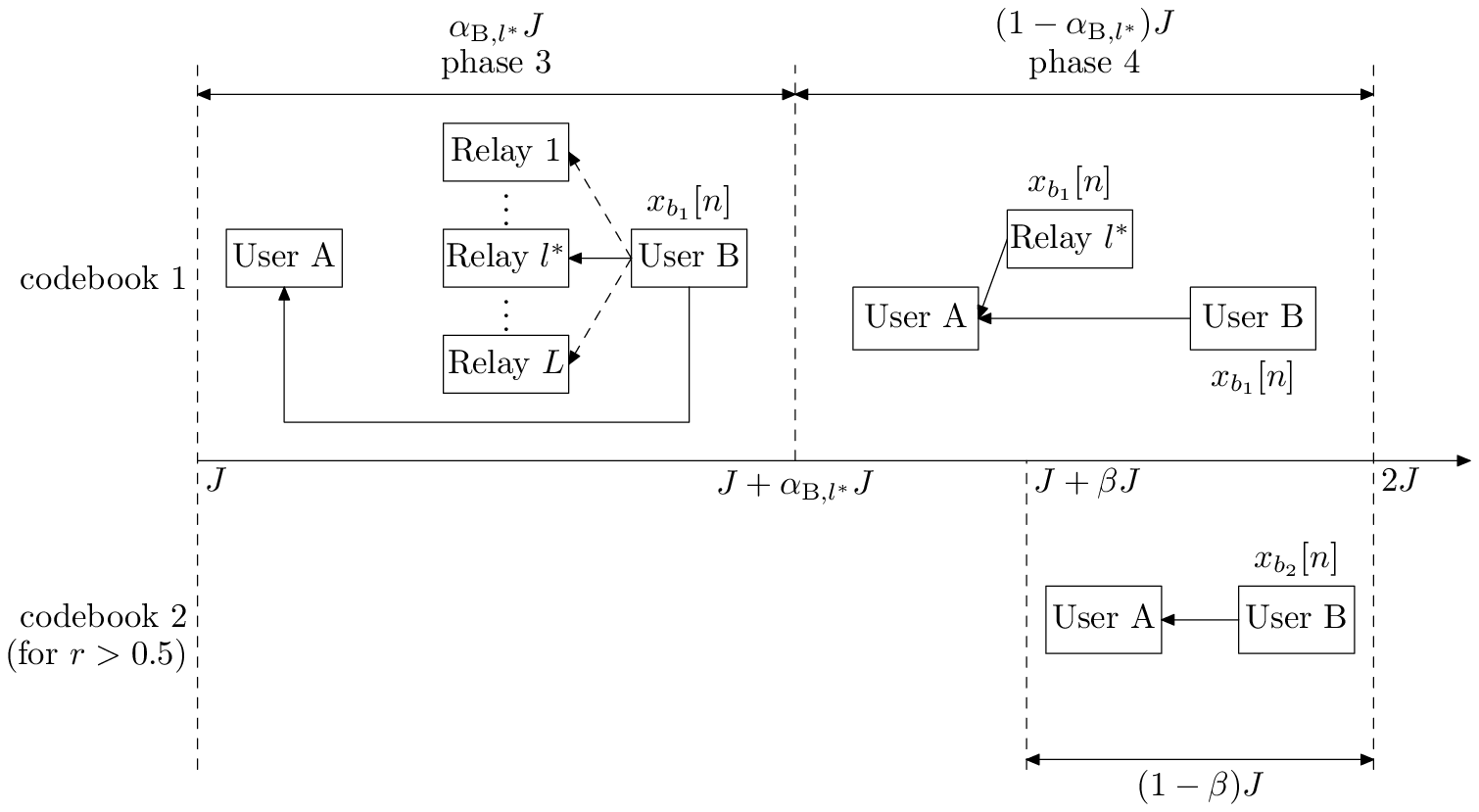}
\caption{Illustration of phase 3 and phase 4 for the RS-EDDF\&NC scheme.}
\label{fig:RS_EDDF_NC_phases_3_and_4}
\end{figure}

In phase 4, relay $l^*$ re-encodes the message by using the codebook 1 and
cooperatively transmits the corresponding part of the codeword to the destination (user A) in the remaining intervals if the outage did not occur. Also, the user B uses codebook 2 to transmit an independent codeword $x_{b_2}[n]$ during the interval $(J + \beta J, 2J]$. The received signals at the destination (user A) in this phase can be expressed as
\begin{equation}
\yAn =
\begin{cases}
\hBA x_{b_1}[n] + \hlstarA x_{b_1}[n] + \wAn, & J + \alphaBlstar J < n \leq J + \beta J \\
\hBA x_{b_1}[n] + \hlstarA x_{b_1}[n] + \hBA x_{b_2}[n] + \wBn, & J + \beta J < n \leq 2J.
\end{cases}
\end{equation}
The system model of phase 3 and phase 4 for the RS-EDDF\&NC is shown in Fig.~\ref{fig:RS_EDDF_NC_phases_3_and_4}.

\subsection{Phase 5 of RS-EDDF\&NC}
\label{sec:Phase 5 of RS-EDDF&NC}

In phase 5, both RS-EDDF\&NC and RS-DDF\&NC have the same system
procedure as described in section \ref{sec:Phase 5 of RS-DDF&NC}. Relay $l^*$ performs exclusive-or (XOR) on the two decoded messages from user A and user B and encodes it by using codebook 1. The received signals at user A and user B are given in (\ref{eq:y_A[n]5}) and (\ref{eq:y_B[n]5}), respectively. The system model is shown in Fig.~\ref{fig:RS_DDF_NC_phase_5}.

\subsection{Outage Probability of RS-EDDF\&NC}
\label{sec:Outage Probability of RS-EDDF&NC}

Following the analysis as in section \ref{sec:Outage Probability of RS-DDF&NC}, the outage event for user A and user B are
\begin{equation}
\label{eq:OAEDDFNC}
\OAEDDFNC = \OBAEDDFNC \cap \OlstarAEDDFNC
\end{equation}
and
\begin{equation}
\label{eq:OBEDDFNC}
\OBEDDFNC = \OABEDDFNC \cap \OlstarBEDDFNC,
\end{equation}
respectively, with
\begin{equation}
\label{eq:OABEDDFNC}
\OABEDDFNC = O_1 \cup O_2 \cup O_3,
\end{equation}
where
\begin{equation}
\label{eq:O_1}
O_1 = \left\{ \frac{2}{3} \log_2 \left( 1 + \rho \hABs \right) < R_1 \right\},
\end{equation}
\begin{equation}
\label{eq:O_2}
O_2 = \left\{ \frac{2(1 - \beta)}{3} \log_2 \left( 1 + \rho \hABs \right) < R_2 \right\},
\end{equation}
and
\begin{equation}
\label{eq:O_3}
O_3 = \left\{ \frac{2 \beta}{3} \log_2 \left( 1 + \rho \hABs \right) + \frac{2(1 - \beta)}{3} \log_2 \left( 1 + 2 \rho \hABs \right) < R \right\}.
\end{equation}
Here, $O_1$ is the outage event corresponds to rate $R_1$, $O_2$ is the outage event corresponds to rate $R_2$ and $O_3$ is the outage event corresponds to the total rate $R$. In addition, $\OBAEDDFNC = \OABEDDFNC$ and $\rho$ is the effective SNR given by (\ref{eq:rho}).

Moreover, the outage event between relay $l^*$ and user A is
\begin{equation}
\label{eq:OlstarAEDDFNC}
\OlstarAEDDFNC = \EoneEDDFNC \cup \ElstarAEDDFNC,
\end{equation}
where
\begin{equation}
\EoneEDDFNC = \left\{ \beta < \alphaBlstar \right\}
\end{equation}
and
\begin{equation}
\label{eq:ElstarAEDDFNC}
\ElstarAEDDFNC = \left\{ \frac{2 + \alphaAlstar - \alphaBlstar}{3} \log_2 \left( 1 + \rho \hlstarAs \right) < R_1 \right\}.
\end{equation}

Similarly, the outage event between relay $l^*$ and user B is
\begin{equation}
\label{eq:OlstarBEDDFNC}
\OlstarBEDDFNC = \EtwoEDDFNC \cup \ElstarBEDDFNC,
\end{equation}
where
\begin{equation}
\label{eq:EtwoEDDFNC}
\EtwoEDDFNC = \left\{ \beta < \alphaAlstar \right\}
\end{equation}
and
\begin{equation}
\label{eq:ElstarBEDDFNC}
\ElstarBEDDFNC = \left\{ \frac{2 + \alphaBlstar - \alphaAlstar}{3} \log_2 \left( 1 + \rho \hlstarBs \right) < R_1 \right\}.
\end{equation}

The outage event of the overall system for RS-EDDF\&NC can be written as
\begin{align}
\label{eq:ORSEDDFNC}
\ORSEDDFNC
& = \OAEDDFNC \cup \OBEDDFNC \notag \\
& = (\OBAEDDFNC \cap \OlstarAEDDFNC) \cup (\OABEDDFNC \cap \OlstarBEDDFNC) \notag \\
& = \OABEDDFNC \cap (\OlstarAEDDFNC \cup \OlstarBEDDFNC) \notag \\
& = [O_1 \cup O_2 \cup O_3] \cap [(\EoneEDDFNC \cup \ElstarAEDDFNC) \cup (\EtwoEDDFNC \cup \ElstarBEDDFNC)]
\end{align}
Let $Z_1 = \hAlstars$ and $Z_2 = \hBlstars$, the outage probability of RS-EDDF\&NC can be expressed as
\begin{equation}
\label{eq:PoutRSEDDFNC}
\PoutRSEDDFNC = \int_0^\infty \int_0^\infty \P[\ORSEDDFNC] f_{Z_1, Z_2}(z_1, z_2) dz_1 dz_2.
\end{equation}

The simulation parameters are listed in Table \ref{tab:simulation_parameters}.

\begin{table}[!h]
\centering
\caption{Simulation parameters}
\label{tab:simulation_parameters}
\begin{tabular}{lll}
\hline		
\textbf{Symbol} & \textbf{Meaning} &  \textbf{Value} \\ \hline
$J$ & number of symbol intervals in a codeword & 10 \\
$L$ & number of relays & 1, 2, 3, 4, 6, 8 \\
$R$ & data rate & 1, 2, 3, 4, 5 bps/Hz \\
$\rho$ & signal-to-noise ratio & 1--30 dB		
\\ \hline
\end{tabular}
\end{table}

\section{Numerical Results}
\label{sec:Numerical Results}

\subsection{Rayleigh Fading Channel}
\label{sec:Rayleigh Fading Channel}

\begin{figure}
\centering
\includegraphics{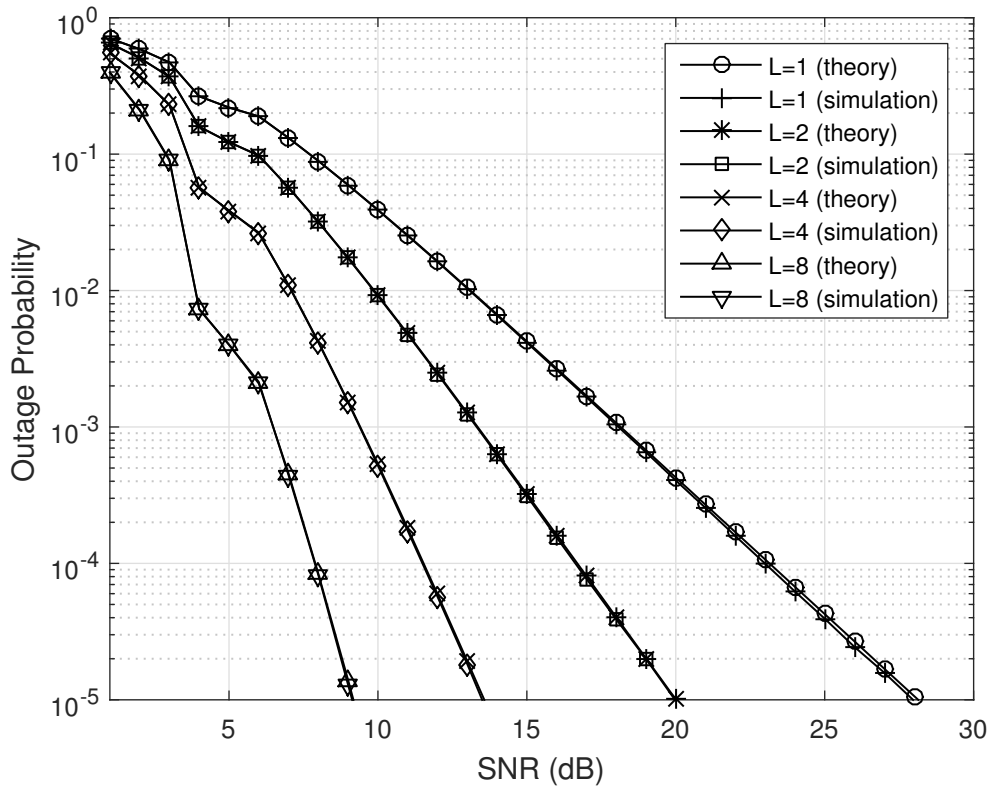}
\caption{Outage probabilities of RS-EDDF\&NC with various numbers of relays and transmission rate $R = 1$ (bps/Hz) under Rayleigh fading channel.}
\label{fig:Pout_RS_EDDF_NC_R1_Rayleigh}
\end{figure}

\begin{figure}
\centering
\includegraphics{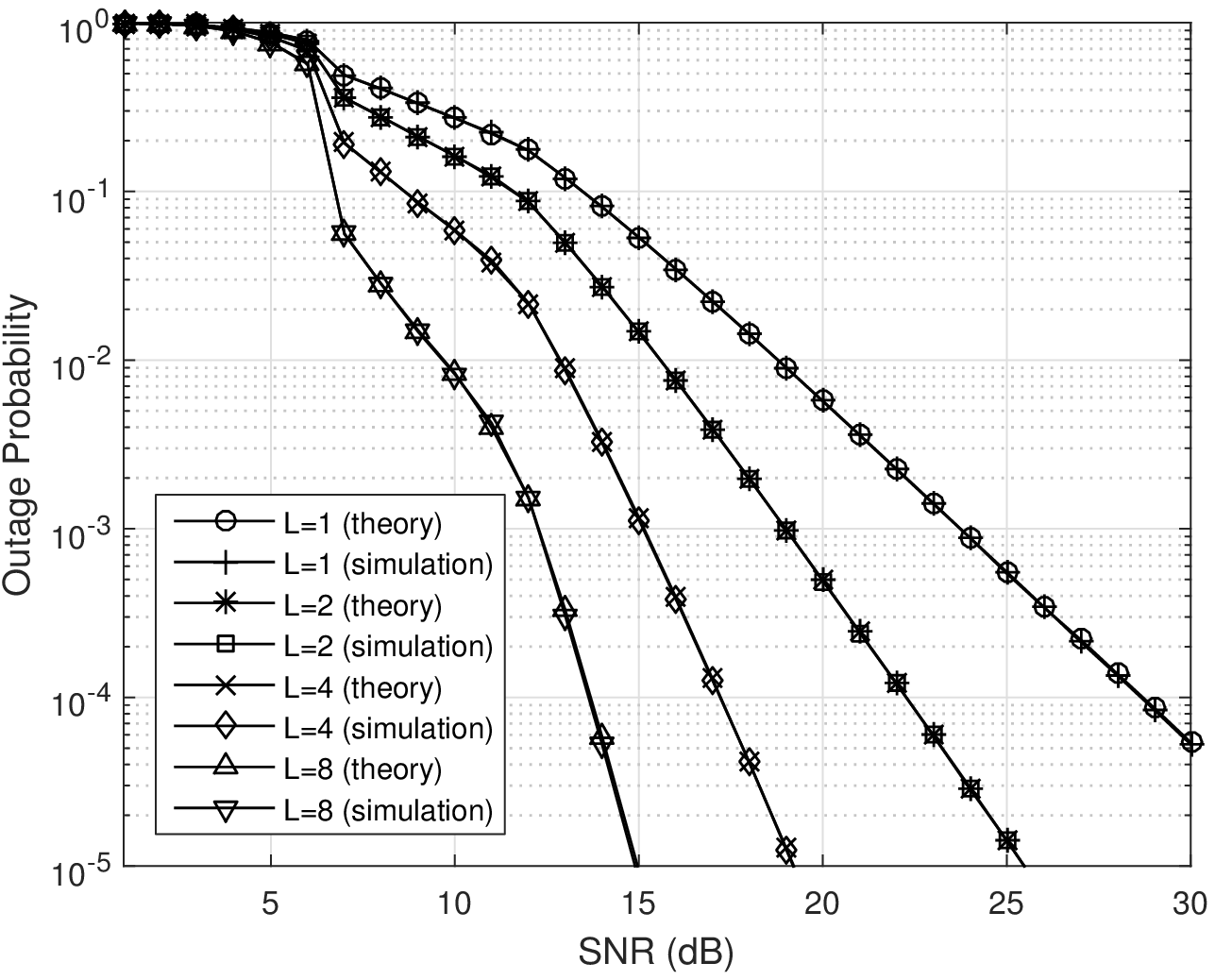}
\caption{Outage probabilities of RS-EDDF\&NC with various numbers of relays and transmission rate $R = 2$ (bps/Hz) under Rayleigh fading channel.}
\label{fig:Pout_RS_EDDF_NC_R2_Rayleigh}
\end{figure}

\begin{figure}
\centering
\includegraphics{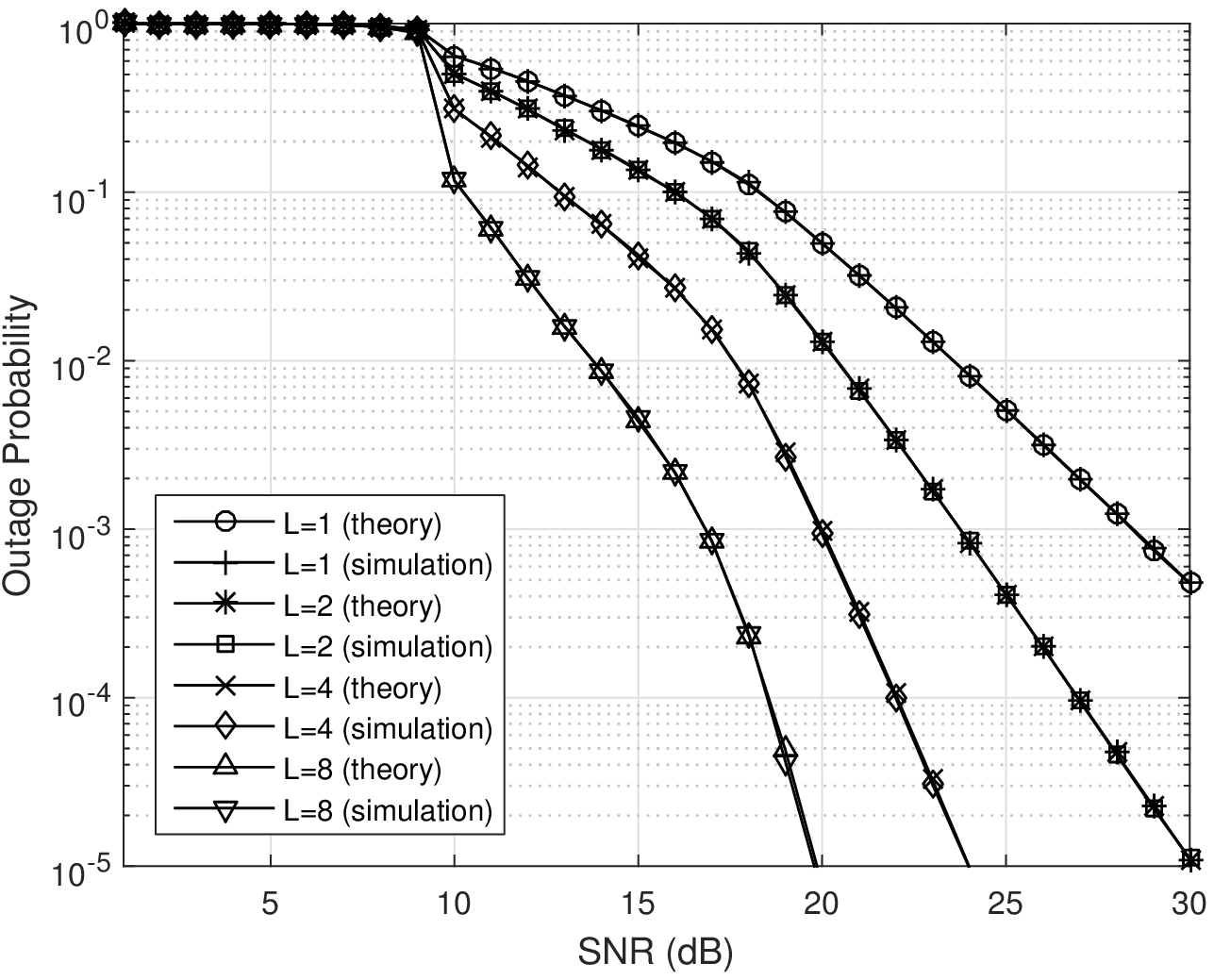}
\caption{Outage probabilities of RS-EDDF\&NC with various numbers of relays and transmission rate $R = 3$ (bps/Hz) under Rayleigh fading channel.}
\label{fig:Pout_RS_EDDF_NC_R3_Rayleigh}
\end{figure}

\begin{figure}
\centering
\includegraphics{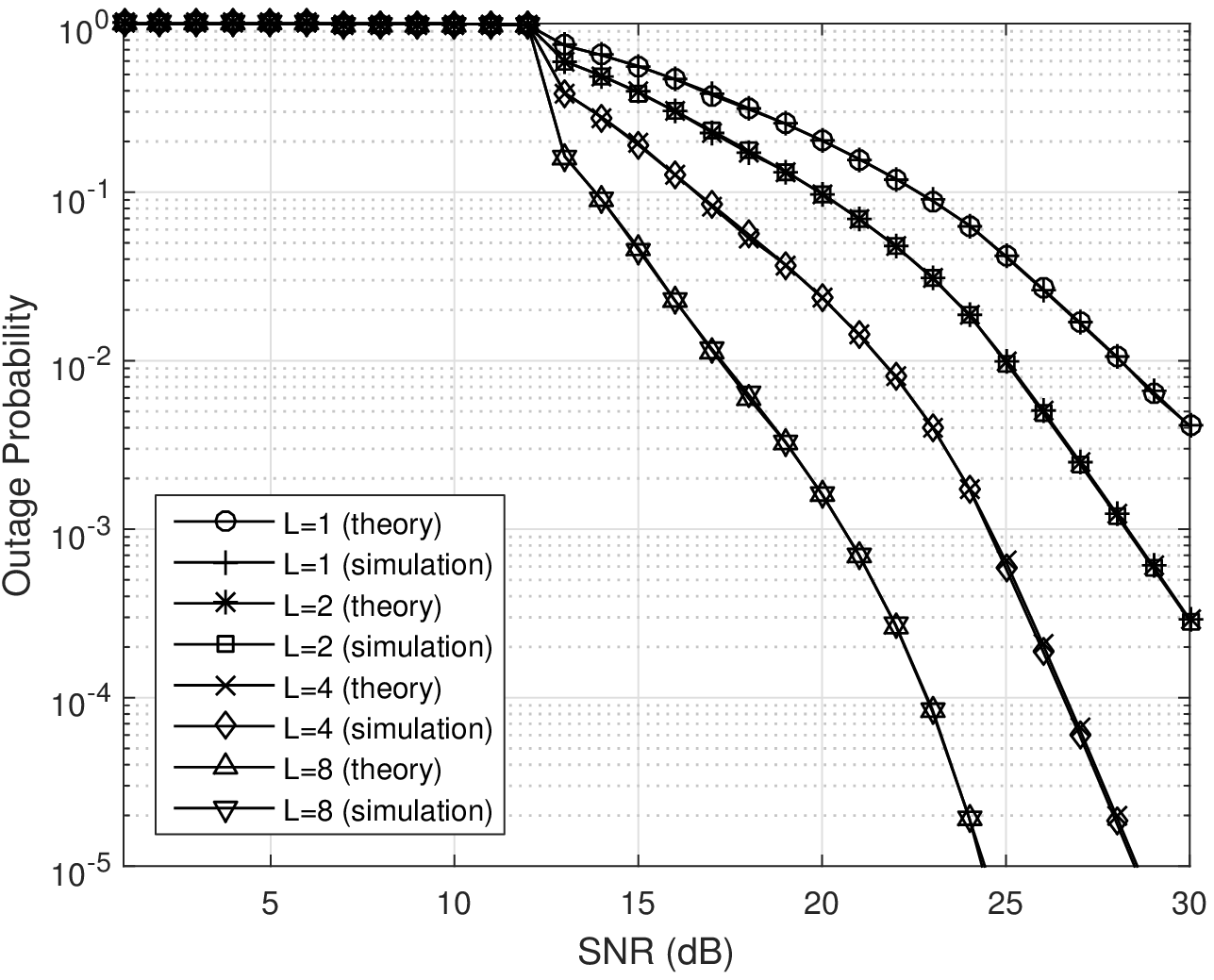}
\caption{Outage probabilities of RS-EDDF\&NC with various numbers of relays and transmission rate $R = 4$ (bps/Hz) under Rayleigh fading channel.}
\label{fig:Pout_RS_EDDF_NC_R4_Rayleigh}
\end{figure}

\begin{figure}
\centering
\includegraphics{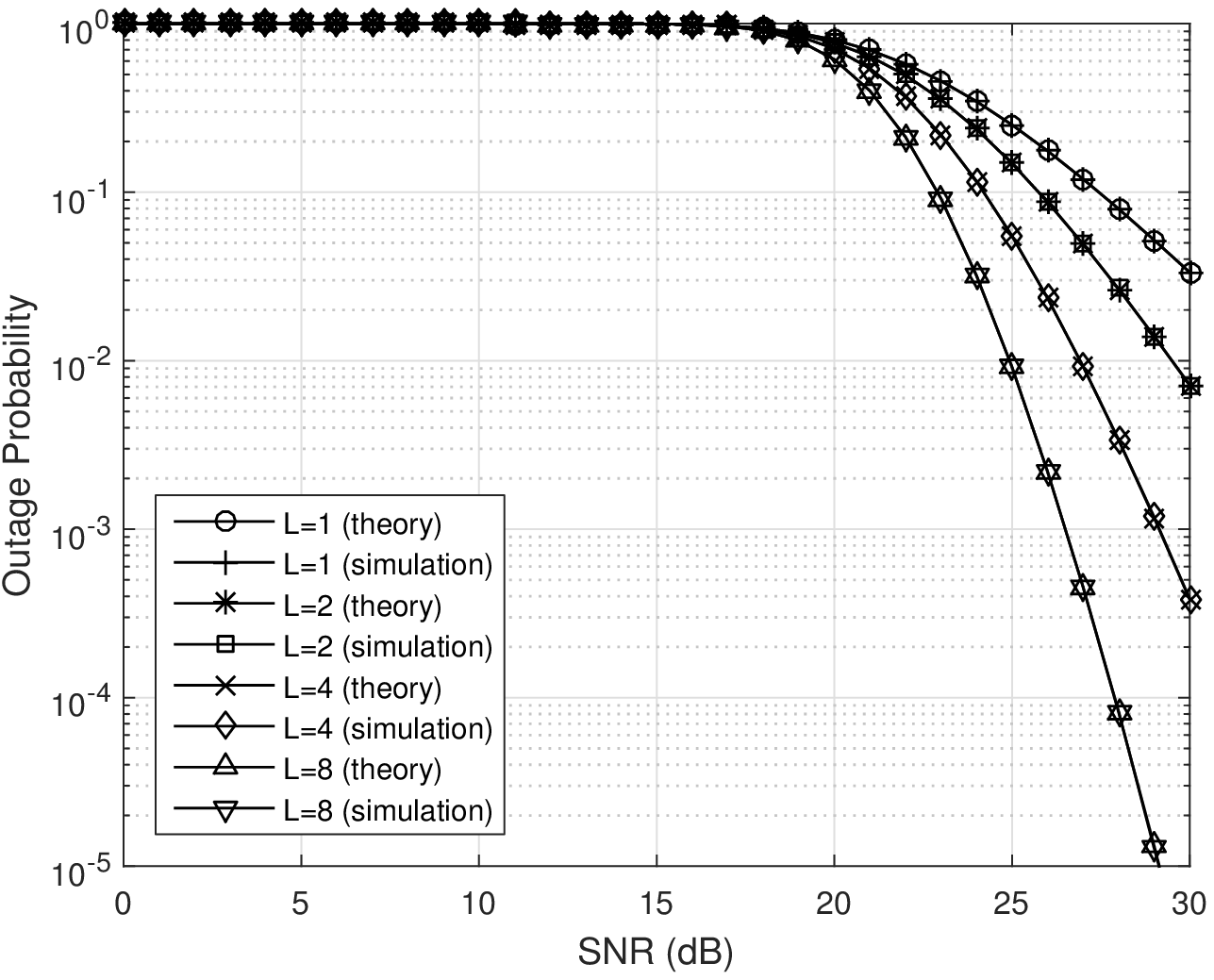}
\caption{Outage probabilities of RS-EDDF\&NC with various numbers of relays and transmission rate $R = 5$ (bps/Hz) under Rayleigh fading channel.}
\label{fig:Pout_RS_EDDF_NC_R5_Rayleigh}
\end{figure}

Figs.~\ref{fig:Pout_RS_EDDF_NC_R1_Rayleigh}--\ref{fig:Pout_RS_EDDF_NC_R5_Rayleigh} show comparisons of outage probability for various numbers of relays under i.i.d.~Rayleigh fading channel with different transmission rates. In this case, channel coefficients are i.i.d.~circularly symmetric complex Gaussian random variables with zero mean and unit variance, i.e., $h_{X,Y} \stackrel{i.i.d.}{\sim} {\cal CN}(0, 1)$. In these figures, one can see that curves descend rapidly in specific SNR regions. These regions correspond to the multiplexing gain $r \in (0.5, 1]$. Curves in these regions are outage probabilities for RS-EDDF\&NC. While multiplexing gain $r \notin (0.5, 1]$, RS-EDDF\&NC is identical to RS-DDF\&NC. As expected, more available relays results in smaller outage probability, and higher transmission rate causes larger outage probability.

\subsection{Comparisons}
\label{sec:Comparisons}

\begin{figure}
\centering
\includegraphics{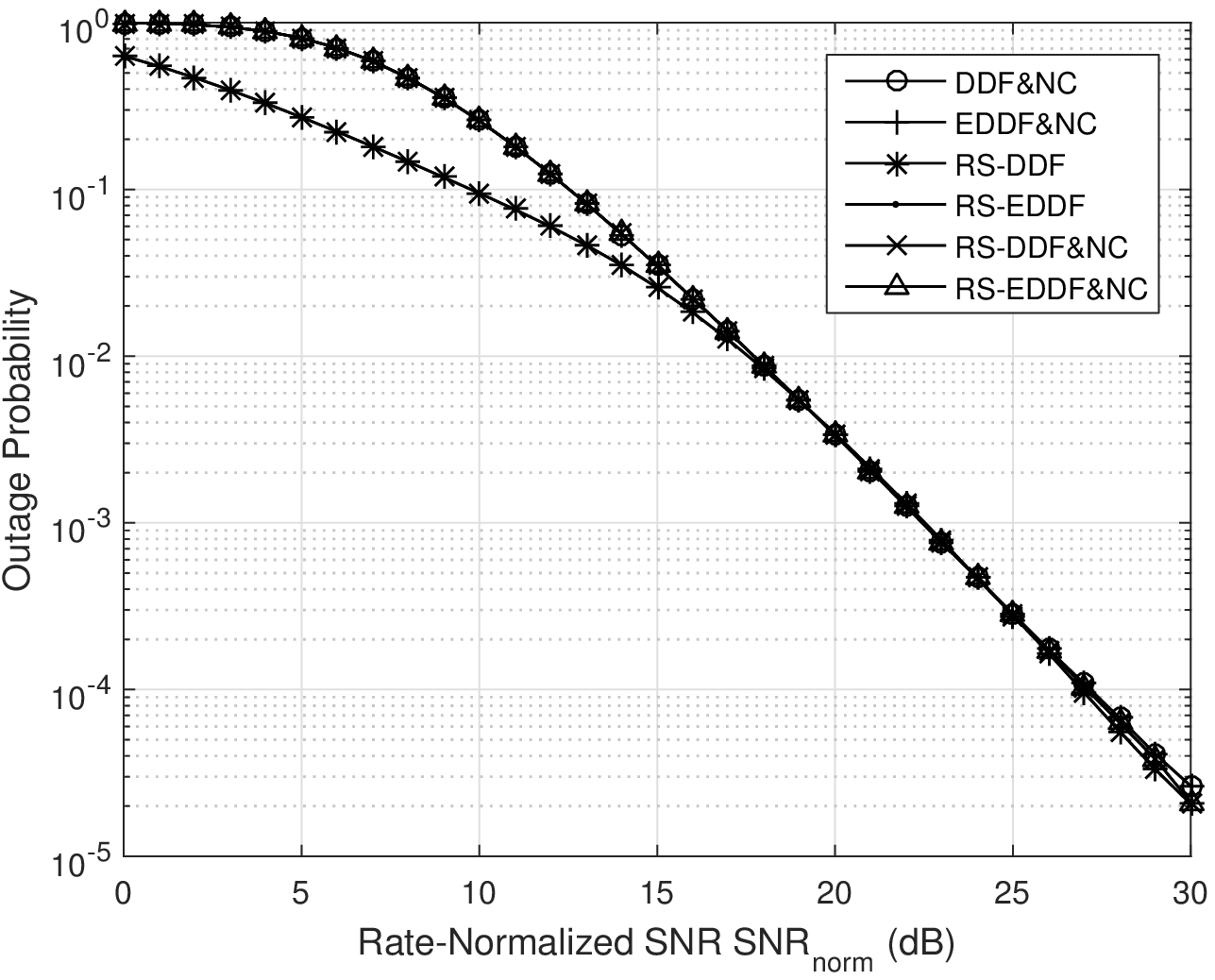}
\caption{Comparison of outage probability for various decode-and-forward relaying schemes with number of relays $L = 1$ and transmission rate $R = 5$ (bps/Hz) under Rayleigh fading channel.}
\label{fig:comparisonL1R5}
\end{figure}

\begin{figure}
\centering
\includegraphics{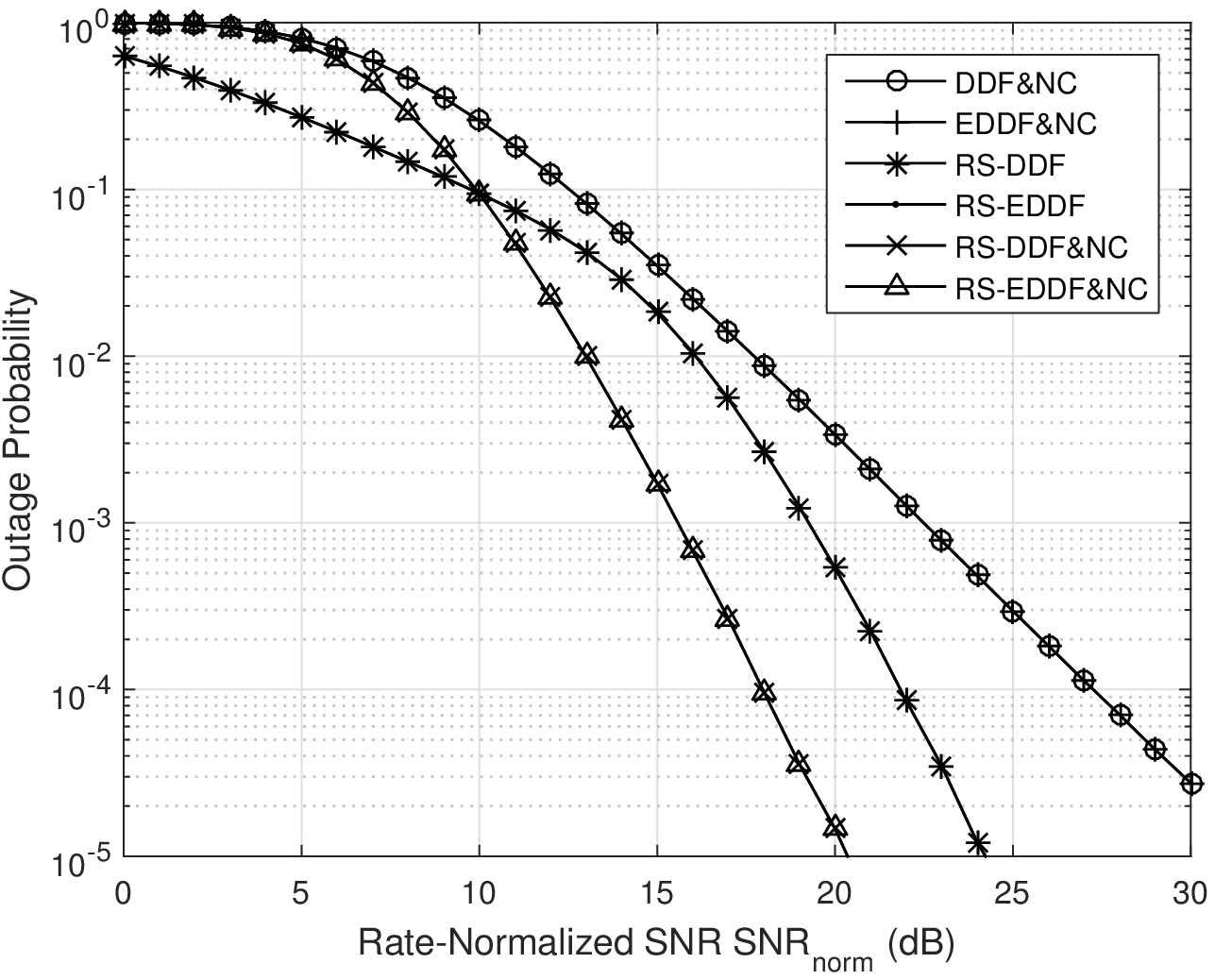}
\caption{Comparison of outage probability for various decode-and-forward relaying schemes with number of relays $L = 3$ and transmission rate $R = 5$ (bps/Hz) under Rayleigh fading channel.}
\label{fig:comparisonL3R5}
\end{figure}

\begin{figure}
\centering
\includegraphics{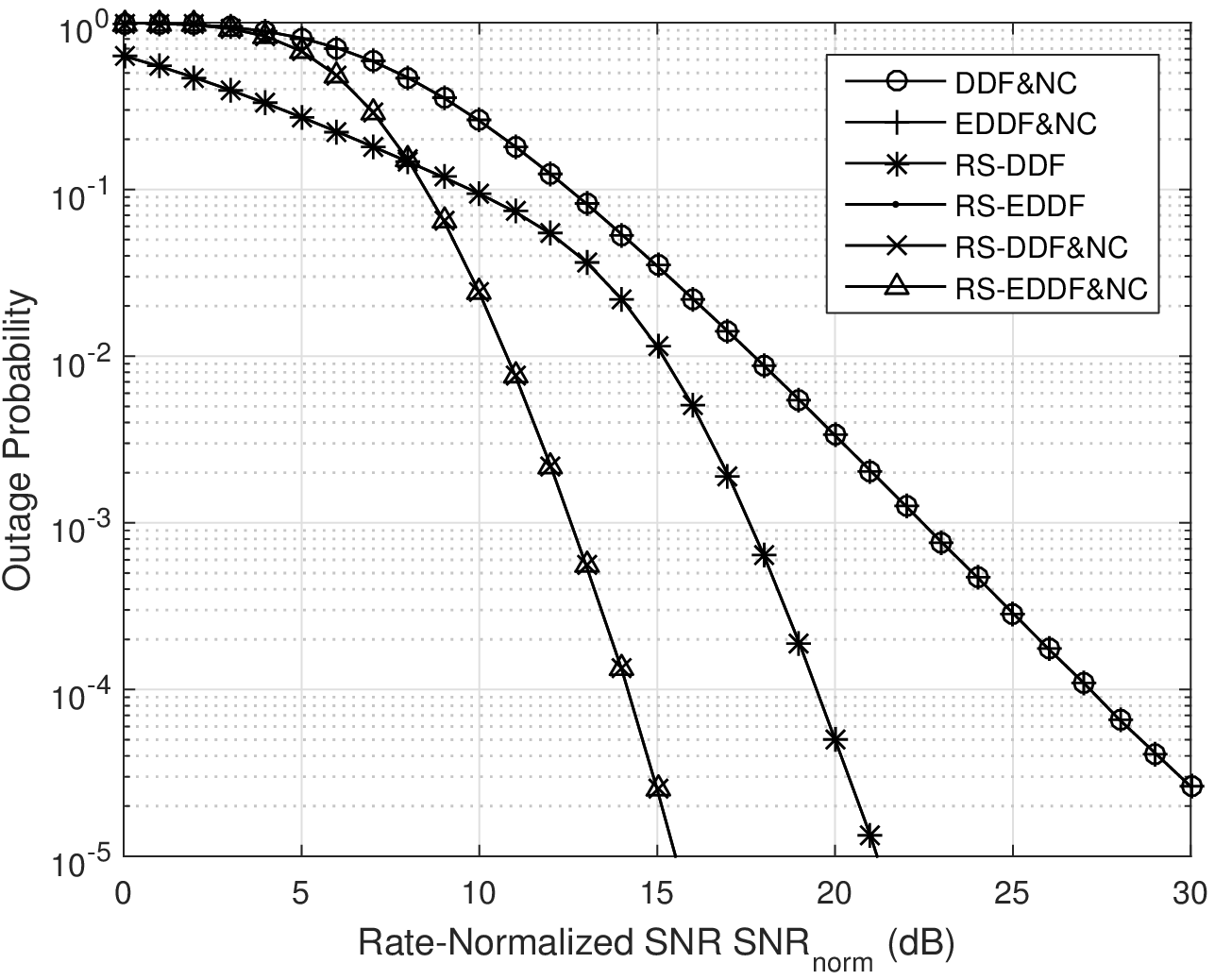}
\caption{Comparison of outage probability for various decode-and-forward relaying schemes with number of relays $L = 6$ and transmission rate $R = 5$ (bps/Hz) under Rayleigh fading channel.}
\label{fig:comparisonL6R5}
\end{figure}

Figs.~\ref{fig:comparisonL1R5}--\ref{fig:comparisonL6R5} compare the outage probability of the RS-EDDF\&NC scheme with other considered decode-and-forward schemes under Rayleigh fading channel. Because different designs may support different data rates, in order to compare these schemes fairly, we plot the outage probability scaled with rate-normalized SNR \cite{Forney1998} defined as
\begin{equation}
\SNRnorm=\frac{\SNR}{2^R-1},
\end{equation}
which is normalized by the minimum SNR required to achieve spectral efficiency $R$. Here, all schemes adopt the same relay selection criterion given by (\ref{eq:l^*}) except DDF\&NC and EDDF\&NC. The DDF\&NC and EDDF\&NC schemes select a relay randomly. The RS-DDF scheme has the same system model in section \ref{sec:RS-DDF&NC} except the phase 5. Thus, the RS-DDF only has phases 1-4 and the whole system duration is only $2J$. Similarly, the RS-EDDF also has the same system model except the phase 5 in section \ref{sec:RS-EDDF&NC}. Further, if multiplexing gain $r \leq 0.5,$ the RS-EDDF scheme is identical to the RS-DDF scheme. From Fig.~\ref{fig:comparisonL1R5}, we can see that RS-DDF and RS-EDDF outperform other schemes that combine the network coding technique. This is due to the benefit of the network coding is degraded since the system duration of both RS-DDF\&NC and RS-EDDF\&NC is longer. In contrast, the network coding gain influentially lowers the outage probability as $L$ increases. Figs.~\ref{fig:comparisonL3R5}--\ref{fig:comparisonL6R5} reveal this
result. Also, we can see that the performance loss is large if we select a relay randomly. This shows the importance of relay selection strategies.

\section{Conclusions}
\label{sec:Conclusions}

In this paper, we investigated the performance of EDDF adopted the relay selection protocol and network coding techniques. We analyzed the outage probabilities of RS-DDF\&NC and RS-EDDF\&NC and demonstrated the system performance under Rayleigh fading channel. We further compared the RS-EDDF\&NC scheme with other considered decode-and-forward relaying schemes.

According to the results, we found that RS-DDF and RS-EDDF without network coding outperform RS-DDF\&NC and RS-EDDF\&NC in a two-way network without sufficient relays. This is due to the system duration of RS-DDF and RS-EDDF is less than RS-DDF\&NC and RS-EDDF\&NC. If there are sufficient relays, then the NC gain is effective and the RS-EDDF\&NC scheme outperforms all other considered decode-and-forward relaying schemes. In addition, the performance loss is large if we select a relay at random. This shows the importance of relay selection strategies.

\begin{appendices}

\section{Proof of Lemma \ref{lem:fZ1Z2}}
\label{app:fZ1Z2}

Let $L$  be a positive integer. Let
$X_1,\ldots, X_L$, $Y_1,\ldots, Y_L$ be independent and identically
distributed (i.i.d.) exponential random variables with parameter $\lambda=1$.
Then 
\begin{equation}
f_{X_l}(x_l)=\begin{cases}
e^{-x_l},& x_l\geq 0,\\
0, & \textup{otherwise}.
\end{cases}
\end{equation}
\begin{equation}
f_{Y_l}(y_l)=\begin{cases}
e^{-y_l},& y_l\geq 0,\\
0, & \textup{otherwise}.
\end{cases}
\end{equation}

Define
\begin{equation}
S_l=\min\{X_l, Y_l\},\quad
T_l=\max\{X_l, Y_l\}, \quad l=1,\ldots,L
\end{equation}
Then 
\begin{equation}
f_{S_l,T_l}(s_l,t_l)=\begin{cases}
2e^{-s_l-t_l},& 0\leq s_l\leq t_l,\\
0, & \textup{otherwise}.
\end{cases}
\end{equation}
\begin{equation}
f_{S_l}(s_l) = \begin{cases}
2e^{-2s_l},& s_l\geq 0,\\
0, & \textup{otherwise}.
\end{cases}
\end{equation}

\begin{equation}
f_{T_l|S_l}(t_l|s_l)= \begin{cases}
e^{-(t_l-s_l)}, & t_l\geq s_l,\\
0, & \textup{otherwise}.
\end{cases}
\end{equation}

Let
\begin{equation}
l_0 =\arg\max_{l}S_l
\end{equation}
Define
\begin{equation} 
Z_1= X_{l_0},\quad Z_2 = Y_{l_0},\quad
U= \min\{Z_1,Z_2\}, \quad 
V= \max\{Z_1,Z_2\}.
\end{equation} 
Then
\begin{equation}
U=\max\{ S_1,\ldots, S_L\}
\end{equation}
\begin{equation}
F_U(u)= P[U\leq u]= P[S_1\leq u]\cdots P[S_L\leq u]= 
\begin{cases} (1-e^{-2u})^L, & u\geq 0,\\
0, & \textup{otherwise}.
\end{cases}
\end{equation}
\begin{equation}
f_U(u) =  \begin{cases} 2Le^{-2u}(1-e^{-2u})^{L-1}, & u\geq 0,\\
0, & \textup{otherwise}.
\end{cases}
\end{equation}
If $u\geq v\geq 0$ then
\begin{equation}
f_{U,V}(u,v)=f_{V|U}(v|u)f_U(u) = e^{-(v-u)} 2Le^{-2u}(1-e^{-2u})^{L-1} = 2L e^{-(u+v)} (1-e^{-2u})^{L-1}.
\end{equation}
We have
\begin{equation}
f_{U,V}(u,v)=\begin{cases}
2L e^{-(u+v)} (1-e^{-2u})^{L-1}, & 0\leq u\leq v,\\
0, & \textup{otherwise}. 
\end{cases}
\end{equation}

We also have
\begin{equation}
f_{U,V}(u,v)= \begin{cases}
f_{Z_1,Z_2}(u,v) + f_{Z_1,Z_2}(v,u),& 0\leq u\leq v\\
0, &\textup{otherwise},
\end{cases}
\end{equation}
\begin{equation}
f_{Z_1,Z_2}(a,b)= f_{Z_1,Z_2}(b,a).
\end{equation}

Therefore, 
\begin{equation}
f_{Z_1,Z_2}(z_1,z_2)=\begin{cases}
\frac{1}{2} f_{U,V}(z_1,z_2)= L e^{-(z_1+z_2)} (1-e^{-2z_1})^{L-1}, & 0\leq z_1\leq z_2\\
\frac{1}{2} f_{U,V}(z_2,z_1)= L e^{-(z_1+z_2)} (1-e^{-2z_2})^{L-1}, & 0\leq z_2\leq z_1\\
0, & \textup{otherwise}. 
\end{cases}
\end{equation}

\section{Proof of Theorem \ref{the:PoutRSDDFNC}}
\label{app:PoutRSDDFNC}

First, the probability of event $\OABDDFNC$ is
\begin{align}\label{eq:POABDDFNC}
\POABDDFNC
&=\P\left[\frac{2}{3}\log_2(1+\rho|\hAB|^2)<R\right]\notag\\
&=\P\left[|\hAB|^2<\frac{2^\frac{3R}{2}-1}{\rho}\right]\notag\\
&\stackrel{\rm (a)}{=}1-\exp\left(-\frac{2^{\frac{3R}{2}}-1}{\rho}\right)
\triangleq p_1.
\end{align}
The equality (a) holds because $|\hAB|^2$ is an exponential random variable with rate parameter 1. Moreover,
\begin{align}\label{eq:PEone...}
& \P [ (\EoneDDFNC \cup \ElstarADDFNC) \cup (\EtwoDDFNC \cup \ElstarBDDFNC)] \notag \\
\stackrel{\rm (b)}{=} & \P\left[\left\{\log_2(1+\rho z_2)\leq\frac{JR}{J-1}\right\}
\cup\left\{\frac{2J+\JprimeAlstar-\JprimeBlstar}{3J}\log_2(1+\rho z_1)<R\right\}\right.\notag\\
&\left.\cup\left\{\log_2(1+\rho z_1)\leq\frac{JR}{J-1}\right\}
\cup\left\{\frac{2J+\JprimeBlstar-\JprimeAlstar}{3J}\log_2(1+\rho z_2)<R\right\}\right]\notag\\
\stackrel{\rm (c)}{=}&\P\left[\left\{z_2\leq\frac{2^{JR/(J-1)}-1}{\rho}\right\}\right]\notag\\
&\cup\left\{\left(2J+\min\left\{J,\left\lceil\frac{JR}{\log_2(1+\rho z_1)}\right\rceil\right\}-\min\left\{J,\left\lceil\frac{JR}{\log_2(1+\rho z_2)}\right\rceil\right\}\right)\log_2(1+\rho z_1)<3JR\right\}\notag\\
&\cup\left\{z_1\leq\frac{2^{JR/(J-1)}-1}{\rho}\right\}\notag\\
&\left.\cup\left\{\left(2J+\min\left\{J,\left\lceil\frac{JR}{\log_2(1+\rho z_2)}\right\rceil\right\}-\min\left\{J,\left\lceil\frac{JR}{\log_2(1+\rho z_1)}\right\rceil\right\}\right)\log_2(1+\rho z_2)<3JR\right\}\right]\notag\\
\triangleq&\P[O],
\end{align}
where equality (b) results from (\ref{eq:EoneDDFNC}), (\ref{eq:ElstarADDFNC}), (\ref{eq:EtwoDDFNC}), and (\ref{eq:ElstarBDDFNC}). Equality (c) results from (\ref{eq:JprimeAlstar}) and (\ref{eq:JprimeBlstar}) with the assumption $\rhoAlstar = \rhoBlstar = \rho$. Furthermore, we replace $|\hAlstar|^2$ by $z_1$ and $|\hBlstar|^2$ by $z_2$.

Finally, we can rewrite the outage probability given by (\ref{eq:PoutRSDDFNC}) as
\begin{align}\label{eq:PoutRSDDFNC3}
\PoutRSDDFNC
& = \int_0^\infty \int_0^\infty \P[\ORSDDFNC] f_{Z_1,Z_2}(z_1,z_2)dz_1 dz_2 \notag \\
& = \int_0^\infty \int_0^\infty \P[\OABDDFNC \cap O] f_{Z_1,Z_2} (z_1,z_2) dz_1 dz_2 \notag \\
& \stackrel{\rm (d)}{=} \int_0^\infty \int_0^\infty \P[\OABDDFNC] \P[O] f_{Z_1,Z_2}(z_1,z_2) dz_1 dz_2 \notag \\
& = p_1 \int_0^\infty \int_0^\infty I(O) f_{Z_1,Z_2}(z_1,z_2) dz_1 dz_2,
\end{align}
where equality (d) holds because events $\OABDDFNC$ and $O$ are independent.

\end{appendices}

\section*{Acknowledgment}
The authors would like to thank Professor Chiu-Chu Melissa Liu. She provides the proof of lemma \ref{lem:fZ1Z2}.

\bibliographystyle{IEEEtran}
\bibliography{IEEEabrv,mybibfile}

\end{document}